# House Price Modeling with Digital Census

Enwei Zhu[1,2], Stanislav Sobolevsky[1,*]

[1]Center for Urban Science and Progress, New York University, New York, USA
[2]Hang Lung Center for Real Estate, Tsinghua University, Beijing, China
*Corresponding author. Email: sobolevsky@nyu.edu

**Abstract:** Urban house prices are strongly associated with local socioeconomic factors. In literature, house price modeling is based on socioeconomic variables from traditional census, which is not real-time, dynamic and comprehensive. Inspired by the emerging concept of "digital census"—using large-scale digital records of human activities to measure urban population dynamics and socioeconomic conditions, we introduce three typical datasets, namely 311 complaints, crime complaints and taxi trips, into house price modeling. Based on the individual housing sales data in New York City, we provide comprehensive evidence that these digital census datasets can substantially improve the modeling performances on both house price levels and changes, regardless whether traditional census is included or not. Hence, digital census can serve as both effective alternatives and complements to traditional census for house price modeling.

## 1 Introduction

House price is one of the most important indicators of urban development and progress. A variety of urban stakeholders, including urban planners, policy makers and investors, may benefit from a better modeling of house prices. In real estate studies, it has been extensively documented that the house prices are associated with local socioeconomic factors, such as racial composition [1-3], income level [2, 3], unemployment rate [4], education and school quality [2, 3, 5], age composition [2] and crime level [5, 6]. However, these socioeconomic variables are typically from census data, which have three limitations. First, the release of census data usually lags for at least one year, so they are not available in real time. Second, census data are static within the survey period, so the dynamic information like the intraday human mobility, is obscured. Third, census data may not reflect all the aspects of local socioeconomic conditions that affect house prices. A more real-time, dynamic and comprehensive house price modeling requires new sources for local socioeconomic information.

With the increasing availability of big data in recent years, an emerging body of studies starts to utilize large-scale digital records of human activities as proxies to urban dynamics, including mobile phone connections [7-12], bank card transactions [13-15], geo-tagged Twitter [14, 16-18] and Flickr [14, 19], Wi-Fi [20] and Bluetooth connections [21], taxi usage records [8, 22], as well as 311 service requests [23]. This gives rise to the new concept of "digital census"—measuring urban population dynamics in real time [20, 24]. Moreover, these datasets can deliver deeper insights into local socioeconomic conditions, such as regional structures [11, 25, 26], land use [27, 28], well-being [29], GDP, education level, income level and unemployment rate [23, 30]. In addition, digital census data are theoretically available in real time, providing dynamic and comprehensive information for specific aspects of urban activities. Hence, digital census may serve as alternatives or even provide additional socioeconomic information to traditional census, and overcome the limitations.

In this paper, we introduce digital census into house price modeling. Starting with replicating the models with traditional census variables in real estate literature, we illustrate the extent to which digital census data can reflect local socioeconomic conditions and whether they can serve as effective alternatives to traditional census data. Specifically, we use the house price data at the most granular level—individual property sales, in New York City (NYC) between 2010 and 2015. We then introduce three typical digital census datasets which can cover the same spatiotemporal scope, namely 311 complaints, crime complaints and taxi trips. Intuitively, 311 and crime complaints reflect local issues, while taxi trip is a typical proxy to human mobility, so all of them are able to capture local urban context [23, 28]. In addition, we use traditional census variables from American Community Survey (ACS) and Longitudinal Employer-Household Dynamics (LEHD) as baselines.

Within the hedonic analysis framework, we regress the sale prices on housing characteristics, socioeconomic variables from traditional census and features from digital census. We employ geographically grouped cross validations to ensure that the out-of-sample $R^2$ can represent the model's geographical generalization ability. It shows that digital census can substantially improve the modeling performance, against both baselines with and without traditional census. The results remain robust after imposing regularization or using nonlinear regression techniques. We then model house price changes within the repeat sales framework, and the results show that digital census can help improve the predictive power as well. In most cases, the magnitudes of improvements appear smaller when the traditional census has been incorporated. These results suggest that digital census can provide effective socioeconomic information for modeling house prices, and such information is partially overlapped with but still substantially differs from traditional census. Hence, digital census can serve as both effective alternatives and complements to traditional census for house price modeling.

## 2 Datasets

### 2.1 Housing sales

The housing sales data are from NYC property sales, which are maintained and reported by the New York City Department of Finance. Each sample in this dataset includes the sale price, sale date, unit number and the "BBL" (borough, block and lot number), a unique identification code for property location. Via the "BBL", we match the housing sales to the PLUTO data, which is from the New York Department of City Planning. The PLUTO provides abundant housing characteristics as well as the coordinates for all the properties in NYC. Thus, we can also calculate additional locational attributes, such as the distance to city center and nearest subway station. After data cleaning,[1] our baseline dataset includes 83,876 residential property sales, covering the period between 2010 and 2015.

Following recent real estate studies based on the same dataset [6, 31-35], we use the sale price per unit[2] as the target variable in this study. We also build a set of housing characteristics including both structural and locational attributes[3] as basic variables to explain the sale prices. More details in the housing sales data can be found in Appendix 1.

**2.2 Digital census**

In this study, we use three digital census datasets—311 complaints, crime complaints and taxi-trips. All these datasets are provided as single records, so we should aggregate the records to build features. Specifically, we use a specific spatiotemporal scope—a one-kilometer geographic "circle" and one-year temporal "window" around each property sale—for aggregation when modeling house prices.

**2.2.1 311 complaints**

The 311 service system is implemented by local governments, providing non-emergency municipal services to local citizens and visitors. Via this system, people can complaint and request services for a broad categories of issues, such as noises, illegal parking, dead trees or animals and abandoned vehicles. Hence, these data are likely to contain effective information about the local socioeconomic conditions, since worse conditions are likely to induce more complaints.

NYC Open Data provides all 311 service request records from 2010, which is publicly available and automatically updated daily. Each record includes the time, location and complaint type of the service request. We aggregate the 311 complaint records to three categories of features:

(1) Total complaint volume. The total volume can reflect the general intensities of local complaints.

(2) Complaint type distribution: the ratios of the complaint volumes of different categories to total complaint volume. This vector of relative complaint frequencies reflects the primary reasons of complaints in a given area, which may serve as a signature of urban locations and socioeconomic conditions [23]. There are 120 complaint categories consistently documented from 2010, so only these complaint categories are used, resulting in a 120-dimensional complaint type distribution vector.

---

[1] We first remove the non-residential property sales. We also remove the sales with prices of zero, which are mostly transfers of ownership from parents to children, and then the outliers with lowest 1% and highest 1% sale prices.
[2] Since a transaction may pertain more than one residential unit in the property, the sale price is divided by the unit number in the transaction.
[3] Some other structural attributes are not available in this dataset, for example, the fixtures, materials and the number of rooms, bedrooms and bathrooms.

(3) Complaint timeline: the ratios of the complaint volumes in three-hour bins during a typical week. This 56-dimensional vector reflects relative complaint intensities in different times.

### 2.2.2 Crime complaints

NYC Open Data provides NYPD Complaint Data, including all the felony, misdemeanor, and violation crimes reported to the New York City Police Department (NYPD) from 2006. Similar to 311 complaints, each crime complaint record includes the time, location and complaint type of the crime. Thus, we also build three categories of features:

(1) Total crime volume. The total volume can reflect the general crime rates.

(2) Crime type distribution: same definition as 311 complaints. There are 48 crime categories consistently documented from 2006, resulting in a 48-dimensional crime type distribution vector.

(3) Crime timeline: same definition as 311 complaints.

### 2.2.3 Taxi trips

NYC Taxi and Limousine Commission provides Trip Record Data, including all the taxi trip records in NYC from 2009. Each taxi trip record contains pick-up and drop-off timestamps and coordinates. We aggregate taxi trip records to two categories of features:

(1) Taxi pick-up and drop-off volumes. The total volume can reflect the general outbound and inbound commuter volumes by taxi.

(2) Taxi pick-up and drop-off timelines: same definition as 311 complaints, but distinguish between pick-up and drop-off records, resulting in a 112-dimensional vector.

## 2.3 Traditional census

### 2.3.1 American Community Survey

The American Community Survey (ACS) contains census data on a number of demographic variables. This dataset is updated annually and available at census tract level from 2009. In this study, we select a set of common variables representing the demographic and socioeconomic status of neighborhoods for modeling house prices. Specifically, the selected variables are "population density," "unemployment rate," "median income of families," "poverty rate of families," "ratio of people with bachelor degree," "ratio of people with graduate or higher degree," "ratio of White people," "ratio of African-American people" and "ratio of Asian people."

### 2.3.2 Longitudinal Employer-Household Dynamics

Longitudinal Employer-Household Dynamics (LEHD) provides LEHD Origin-Destination Employment Statistics (LODES), which reports the worker numbers associated with both a home census block and a work census block. In this study, we aggregate worker numbers in each census tract as home and workplace, and thus have two basic but static variables describing the human mobility.

## 2.4 Summary of datasets

Table 1 presents the summary of the datasets and variables used in this study. All the datasets are publicly available. We provide some basic profiles of the digital census in Appendix 2.

Table 1 Summary of datasets and variables

| Dataset | Source (homepage) | Variable category | Variable number |
|---|---|---|---|
| **Property sales** | http://www1.nyc.gov/site/finance/taxes/property-rolling-sales-data.page (Property sales); https://www1.nyc.gov/site/planning/data-maps/open-data.page (PLUTO) | Sale price per unit | 1 |
| | | Basic housing characteristics | 30 |
| **311 complaints** | https://data.cityofnewyork.us/Social-Services/311-Service-Requests-from-2010-to-Present/erm2-nwe9 (311 Service Requests) | Complaint volume | 1 |
| | | Complaint type distribution | 120 |
| | | Complaint timeline | 56 |
| **Crime complaints** | https://data.cityofnewyork.us/Public-Safety/NYPD-Complaint-Data-Historic/qgea-i56i (NYPD Complaint Data Historic) | Crime volume | 1 |
| | | Crime type distribution | 48 |
| | | Crime timeline | 56 |
| **Taxi trips** | http://www.nyc.gov/html/tlc/html/about/trip_record_data.shtml | Taxi pick-up and drop-off volumes | 2 |
| | | Taxi pick-up and drop-off timelines | 112 |
| **ACS** | https://www.census.gov/programs-surveys/acs/ | Socioeconomic variables | 9 |
| **LEHD** | https://lehd.ces.census.gov/data/ | Worker numbers as residence and workplace areas | 2 |

# 3 Modeling house prices

In this section, we model individual housing sale prices within the hedonic analysis framework. We notice that the hedonic models in real estate literature typically include internal area dummies. However, practical applications typically require the models to generalize to unseen areas, which we refer to as "geographical generalization ability", while the area dummies constrain such ability. Hence, we choose to use purely external information, and evaluate models' geographical generalization ability by cross validation grouped by census tracts, zipcodes and community districts. We then examine whether digital census can improve model performances, and check whether the improvements are robust across different regression methods.

## 3.1 Specification and experiments

A hedonic model is typically specified in a semi-log form, regressing the log sale price on a vector of housing characteristics and time dummies. Additionally, neighborhood socioeconomic factors have important effects on local house prices [1-6], so they should also be incorporated in to the hedonic models, if available. As outlined previously, we introduce socioeconomic information from both traditional and digital censuses. Therefore, the complete modeling specification is:

$$log(P_{it}) = f(HC_{it}, D_t, TC_{it}, DC_{it}) + \varepsilon_{it}$$

where $P_{it}$ is the per unit sale price of house *i* at time *t*; *HC* is a vector of housing characteristics; *D* is a vector of quarterly (seasonal) dummies; *TC* is a vector of socioeconomic variables from traditional census; *DC* is a vector of digital census features; $f(\cdot)$ is a function to fit; and $\varepsilon$ is an *i.i.d.* noise term.

However, the release of traditional census data typically lags for one year or more, so they are not reliable for real-time applications. Hence, we separately build two baselines, one without traditional census to test whether digital census can replace traditional census to capture local socioeconomic conditions, and the other with traditional census to examine whether digital census can provide additional socioeconomic information to traditional census.

We assess the model performance by the average $R^2$ values, which are calculated by five-fold cross validations repeated for 20 times. We typically report both the average $R^2$ values and the corresponding standard deviations.

**3.2 Modeling with internal area dummies**

In real estate studies, hedonic model is an important tool to identify the effects of particular factors on house prices. In most cases, the real estate researchers pay most attention to the consistency of coefficient estimation, so they would choose to include control variables as many as possible to avoid the potential omitted variable problem. Hence, they typically introduce a large set of internal area dummies to control the unobservable heterogeneities among different locations. For example, in the existing studies based on the same housing sales dataset, the researchers include the census tract dummies and quarter dummies, namely the "tract dummies + quarter dummies", or even their interaction terms, namely the "tract*quarter dummies" [6, 31-35].[4] In this manner, coefficient estimation tends to be relatively consistent, accompanied with high in-sample $R^2$ values ranging from about 0.78 to 0.85.

We start from replicating their modeling results on our dataset. We firstly model log sale price with tract*quarter dummies as well as basic housing characteristics. Panel A of Table 2 presents the modeling results. This model reaches an extremely high in-sample $R^2$ as 0.7823, which is located within the range of $R^2$ values reported by related real estate studies. However, it has a relatively low out-of-sample $R^2$ as 0.4405, suggesting that its generalization ability is not very satisfactory even within same tract-quarter group. Moreover, averagely 15.09% of the testing samples are beyond the coverage of tract*quarter dummies of corresponding training samples, so these testing samples cannot be predicted by the fitted model. Both traditional and digital census variables cannot enter this model, since the tract*quarter dummies have explained all the potential variance between the tract-quarter groups.

---

[4] Assume there are *N* census tracts and *T* quarters. "Tract dummies + quarter dummies" means that *N-1* dummies are constructed for different census tracts and *T-1* dummies are constructed for different quarters, so there are totally *N+T-2* dummies. "Tract*quarter dummies" means that *N\*T-1* dummies are constructed for different combinations of census tracts and quarters. The former assumes the orthogonality between tract dummies and quarter dummies, while the later does not.

We then replace the tract*quarter dummies with tract dummies + quarter dummies, and in this case we can introduce traditional census socioeconomic variables and digital census features. Panels B and C present the modeling results for scenarios with and without traditional census, respectively. First, before introducing traditional and digital censuses, the in-sample $R^2$ decreases to 0.6214, while the out-of-sample $R^2$ increases to 0.5961, which means a large improvement of generalization ability compared with the specification with tract*quarter dummies. Additionally, only 0.07% of the testing samples are unpredictable because of not being covered by training samples. Second, 311 complaints, crime complaints and taxi trip features can separately improve both the in-sample and out-of-sample $R^2$ values, against baselines with or without socioeconomic variables. Third, combination of all three digital census datasets can achieve better performances than any single dataset. These results suggest that digital census can provide additional socioeconomic information for house price modeling even with district dummies incorporated, although we also admit that the magnitudes of improvements are very small, which seems to have limited contribution to practical applications.

Table 2 Average $R^2$ values for hedonic models with district dummies

| | No digital census | 311 complaint | Crime complaint | Taxi trips | All digital census |
|---|---|---|---|---|---|
| **Panel A: With tract*quarter dummies** | | | | | |
| **In-sample $R^2$** | 0.7823 (0.0004) | - | - | - | - |
| **Out-of-sample $R^2$** | 0.4405 (0.0025) | - | - | - | - |
| **Testing missing rate** | 0.1509 | - | - | - | - |
| **Panel B: With tract dummies + quarter dummies, without traditional census** | | | | | |
| **In-sample $R^2$** | 0.6214 (0.0004) | 0.6258 (0.0004) | 0.6230 (0.0004) | 0.6253 (0.0004) | 0.6293 (0.0004) |
| **Out-of-sample $R^2$** | 0.5961 (0.0017) | 0.5986 (0.0017) | 0.5965 (0.0017) | 0.5990 (0.0017) | 0.6000 (0.0017) |
| **Testing missing rate** | 0.0007 | 0.0007 | 0.0007 | 0.0007 | 0.0007 |
| **Panel C: With tract dummies + quarter dummies, with traditional census** | | | | | |
| **In-sample $R^2$** | 0.6236 (0.0004) | 0.6269 (0.0004) | 0.6249 (0.0004) | 0.6265 (0.0004) | 0.6300 (0.0004) |
| **Out-of-sample $R^2$** | 0.5982 (0.0017) | 0.5997 (0.0017) | 0.5983 (0.0017) | 0.6001 (0.0017) | 0.6006 (0.0017) |
| **Testing missing rate** | 0.0007 | 0.0007 | 0.0007 | 0.0007 | 0.0007 |

Notes: (1) All columns incorporate basic housing characteristics by default. (2) The average $R^2$ values are estimated by five-fold cross validation repeated for 20 times. (3) The corresponding standard deviations are reported in the

parentheses. (4) Testing missing rate is the ratio of testing samples that are beyond tract coverage of training samples, and consequently unable to be predicted.

However, for the objective of predictive power instead of coefficient estimation, the district-dummy-based approach itself is not preferable in practical applications since these dummies are internal information. More specifically, they are not applicable to areas beyond the data coverage, so have no geographical generalization ability. As mentioned above, even random train-test splits would result in a part of unpredictable testing samples. While in practice, it is a common case to fit models in an area but apply them to other areas, for example, estimating the house prices for areas with limited transactions, or even for vacant lands as investment implications for future development. Hence, we should only rely on the external information—those applicable beyond data coverage—to enable models generalize geographically.

### 3.3 Modeling with external information

We exclude the census tract dummies from the models, and use purely external information to model house prices. In this case, the fitted models are able to generalize to the unseen areas. In other words, they are geographically generalizable.

An important consideration is how to evaluate the geographical generalization ability. In standard cross validation, the samples are randomly grouped into training or testing sets. However, we note that the housing prices and characteristics show significant spatial correlation, so the housing sale samples tend to be highly dependent if their locations are close. Such interdependent samples may be separated and grouped into both training and testing sets. As a result, a model can achieve exaggerated out-of-sample prediction performance by fitting "local patterns," which are effective only in a small area but inapplicable to areas beyond the data coverage. In practice, we may care more about "general patterns," which are effective in a large area. Hence, we evaluate geographical generalization ability by the prediction performance in locations beyond the training data.

Specifically, we split the training and testing sets by geographical groups in cross validation. That is, samples from different areas are treated as different groups, and all the samples in the testing set are ensured to come from groups that are not represented in the corresponding training set. We use geographical groups of three granularities: census tracts, zipcodes and community districts. [5] Including the standard one, we totally have four kinds of cross validations.

Table 3 displays the modeling results with purely external information. For standard cross validation, after excluding district dummies, the average out-of-sample $R^2$ values universally decrease across models with different digital or traditional census, which is as expected. However, the improvements by digital census become larger and more significant. Specifically, for the baseline without socioeconomic variables, 311 complaints, crime complaints and taxi trip features can improve $R^2$ values by 0.1150, 0.0881 and 0.0958, respectively; and their combination can improve $R^2$ values by 0.1412. For the baseline with socioeconomic variables, the improvements are smaller, with the best as 0.0275. However, the differences between average $R^2$ values are larger than standard deviations, suggesting the improvements are statistically significant. In addition, the best out-of-sample $R^2$ values achieved by digital census (0.5713 without socioeconomic variables, 0.5820 with socioeconomic variables) are very close to the $R^2$ values by census tract dummies (about 0.60), suggesting that most house price variation between different census tracts have been captured by digital census.

---

[5] The housing sales dataset covers 1,955 census tracts, 169 zipcodes and 60 community districts.

Regarding the three geographically grouped cross validations, the average out-of-sample $R^2$ values universally decrease across different models with the geographical granularities increasing. Coarser granularity means more preference for "general patterns" rather than "local patterns", as well as more differences between training and testing samples, so it is more challenging for out-of-sample prediction. For example, community district has the coarsest granularity, and the corresponding models achieve lowest $R^2$ values. However, the improvements by digital census against both baselines appear slightly larger and still significant, although the standard deviations also increase. These results suggest that digital census can provide substantial information for house price modeling, and more importantly, the association between digital census features and house prices can effectively generalize to different areas.

Table 3 Average $R^2$ values with cross validation grouped by different geographical granularities

| Cross validation | No digital census | 311 complaint | Crime complaint | Taxi trips | All digital census |
|---|---|---|---|---|---|
| **Panel A: Without traditional census** | | | | | |
| Standard | 0.4300 (0.0017) | 0.5451 (0.0016) | 0.5181 (0.0016) | 0.5258 (0.0016) | 0.5713 (0.0017) |
| Grouped by census tract | 0.4264 (0.0026) | 0.5372 (0.0028) | 0.5141 (0.0027) | 0.5194 (0.0028) | 0.5624 (0.0028) |
| Grouped by zipcode | 0.4093 (0.0058) | 0.5138 (0.0070) | 0.4938 (0.0073) | 0.5039 (0.0070) | 0.5454 (0.0058) |
| Grouped by community district | 0.3460 (0.0135) | 0.4782 (0.0077) | 0.4566 (0.0092) | 0.4495 (0.0100) | 0.5065 (0.0077) |
| **Panel B: With traditional census** | | | | | |
| Standard | 0.5545 (0.0017) | 0.5697 (0.0017) | 0.5611 (0.0017) | 0.5708 (0.0016) | 0.5820 (0.0016) |
| Grouped by census tract | 0.5296 (0.0136) | 0.5556 (0.0064) | 0.5495 (0.0068) | 0.5594 (0.0060) | 0.5727 (0.0033) |
| Grouped by zipcode | 0.4949 (0.0272) | 0.5285 (0.0106) | 0.5189 (0.0143) | 0.5410 (0.0077) | 0.5575 (0.0056) |
| Grouped by community district | 0.4864 (0.0126) | 0.5066 (0.0091) | 0.5010 (0.0096) | 0.5097 (0.0087) | 0.5163 (0.0095) |

Notes: (1) All columns incorporate basic housing characteristics and quarter dummies by default. (2) The average $R^2$ values are out-of-sample, estimated by five-fold cross validation repeated for 20 times. (3) The corresponding standard deviations are reported in the parentheses.

## 3.4 Regularized and nonlinear regression methods

All the models above are fitted by linear regression, so we introduce some machine learning modeling techniques and examine whether the digital census can still help model house prices. As outlined previously, the numbers of digital census features are relatively large, which is likely to result in over-fitting, so we firstly introduce Lasso [36], which imposes L1-regularization to the linear models. On the other hand, the effects of features on house prices may be nonlinear, so we also train three typical nonlinear regression models, namely the Neural Network [37-39], Random Forest [40, 41] and Gradient Tree Boosting [42, 43].

We conduct feature selection before training each nonlinear model, in order to avoid over-fitting. Specifically, we firstly fit an additional Lasso regression, and only the features with nonzero coefficients are selected to enter the nonlinear model. According to our experiments, the parameter $\alpha$ (L1-penalty) of Lasso would be reasonable when ranging from 1e-4.5 to 1e-3.5. More details in the feature selection procedure can be found in Appendix 3.

We then separately train Lasso, Neural Network, Random Forest and Gradient Tree Boosting and evaluate out-of-sample $R^2$ values by cross validation grouped by zipcodes. We conduct exhaustive hyper-parameter tuning to achieve the best performance for each kind of model. Specifically, for Lasso, we try the parameter $\alpha$ for 1e-6, 1e-5.5, 1e-5, …, 1e-1. For Neural Network, we use the L-BFGS algorithm for optimization [44], and try the hidden unit number for 10, 20, 40, 80, and the L2-penalty term for 0.0005, 0.005, 0.05, 0.5. For Random Forest and Gradient Tree Boosting, we try the tree number for 100, 200, 500, 1000, 2000, the maximum depths of trees for 3, 5 and "all-expanded", and "the number of features considered to expand a node" for the number of all features and its square root and logarithm. For later three models, we also try the parameter $\alpha$ in feature selection for 1e-4.5, 1e-4, 1e-3.5, which is treated as an additional hyper-parameter. In total, we have 11, 48, 135 and 135 sets of hyper-parameters for Lasso, Neural Network, Random Forest and Gradient Tree Boosting, respectively.

For each kind of model, we pick the best set of hyper-parameters and report the corresponding $R^2$ values in Table 4. First, the digital census features are still effective after regularization and nonlinearity introduced. 311 complaints, crime complaints and taxi trips can separately improve the out-of-sample $R^2$ values regardless whether traditional census is included; combining all three digital census datasets can achieve better performances than any single dataset. The best $R^2$ value reaches 0.5803 when using all the datasets and Gradient Tree Boosting. In addition, the improvement magnitudes are similar to those of linear regression. Second, all the four machine learning models can produce higher out-of-sample $R^2$ values than linear regression, and Gradient Tree Boosting can achieve the highest. However, improvements by modeling techniques are substantially smaller than introducing meaningful data, either digital or traditional census.

Table 4 Average $R^2$ values by different regression models

| Regression model | Panel A: Without traditional census | | | | |
|---|---|---|---|---|---|
| | No digital census | 311 complaint | Crime complaint | Taxi trips | All digital census |
| Lasso | 0.4111 (0.0014) | 0.5224 (0.0008) | 0.4994 (0.0009) | 0.5055 (0.0010) | 0.5506 (0.0007) |
| Neural Network | 0.4225 (0.0014) | 0.5320 (0.0010) | 0.5067 (0.0011) | 0.5253 (0.0009) | 0.5554 (0.0007) |
| Random Forest | 0.4291 (0.0017) | 0.5292 (0.0008) | 0.4938 (0.0010) | 0.5057 (0.0010) | 0.5400 (0.0008) |
| Gradient Tree Boosting | 0.4262 (0.0018) | 0.5494 (0.0009) | 0.5141 (0.0010) | 0.5425 (0.0008) | 0.5674 (0.0007) |
| Regression model | Panel B: With traditional census | | | | |
| | No digital census | 311 complaint | Crime complaint | Taxi trips | All digital census |
| Lasso | 0.5087 (0.0056) | 0.5299 (0.0062) | 0.5222 (0.0059) | 0.5391 (0.0048) | 0.5612 (0.0009) |
| Neural Network | 0.5222 (0.0136) | 0.5540 (0.0035) | 0.5359 (0.0132) | 0.5487 (0.0060) | 0.5626 (0.0035) |
| Random Forest | 0.5352 (0.0010) | 0.5472 (0.0009) | 0.5355 (0.0009) | 0.5481 (0.0009) | 0.5518 (0.0008) |
| Gradient Tree Boosting | 0.5615 (0.0009) | 0.5720 (0.0008) | 0.5650 (0.0008) | 0.5767 (0.0008) | 0.5803 (0.0007) |

Notes: (1) All columns incorporate basic housing characteristics and quarter dummies by default. (2) The average $R^2$ values are out-of-sample, estimated by five-fold cross validation (grouped by zipcode) repeated for 20 times. (3) The corresponding standard deviations are reported in the parentheses.

## 4 Modeling house price changes

We note that the spatial variance of sale prices is much larger than the temporal variance in our housing sales data. Additionally, we also place much emphasis on the geographical generalization ability in the previous modeling, so there may arise a concern that the patterns revealed above are only generalizable in the spatial dimension. However, most urban stakeholders may be more interested in potential patterns in the temporal dimension, such as forecasting and interpreting price changes.

In this section, we focus on modeling house price changes, emphasizing the association between digital census and house prices in the temporal dimension. We continue to use individual house as the observation unit, and model the price changes within the repeat sales framework.

Repeat sales is an approach to estimate house price indexes using houses that have been sold at least twice [45, 46]. Simple house price indicators, such as median house prices over periods, have so-called structural problem—the sold houses in different periods have heterogeneous characteristics which affect median house prices as well but do not reflect the price changes for a typical house. Repeat sales method compares the multiple sale prices of a same house, so the estimated price indexes can reflect the price differences independent from housing characteristics, which avoid the structural problem and tend to be accurate.

### 4.1 Specification and experiments

In this study, we model the price changes for repeat sale pairs. Specifically, for each house with at least two sales, we match any two successive sales as a pair, and then calculate the difference in log sale prices for each pair as the target variable. The matched dataset ultimately has 12,052 observations. we then estimate a repeat sales price index using this dataset, which represents the general house price trend for the whole city [45, 46]. For each repeat sale pair, we calculate the differences in sale time and the price index as the most basic predictor variables. We then incorporate the housing characteristics, traditional and digital census features of both sales, and examine whether they can help predict house price changes beyond the general city-level trend. Specifically, for a house $i$ successively sold at times $s$ and $t$, we model the log price difference as:

$$log(P_{it}) - log(P_{is}) = g(t - s, PI_t - PI_s, HC_{is}, HC_{it}, TC_{is}, TC_{it}, DC_{is}, DC_{it}) + \varepsilon_{ist}$$

where $PI$ is the repeat sales price index; $HC$, $TC$ and $DC$ are vectors of housing characteristics, traditional census variables and digital census variables, respectively; $g(\cdot)$ is a function to fit; and $\varepsilon$ is an *i.i.d.* noise term. Similarly, we build two baselines, one with and the other without traditional census; and we access model performance by the average $R^2$ values calculated by five-fold cross validations repeated for 20 times.

Related studies typically model house price changes using median prices or quality-constant price indexes, such as repeat sales index [23, 47, 48]. In addition to the structural problem for median prices as described above, those price indicators are firstly estimated for specific regions so the corresponding price changes in are indirect. However, the changes in repeat sale prices are based on the observed price differences for individual houses. We argue that modeling such price changes is more useful for practical implications.

**4.2 Modeling results**

Similarly, we train four machine learning models—Lasso, Neural Network, Random Forest and Gradient Tree Boosting, and evaluate out-of-sample $R^2$ values by cross validation grouped by zipcodes. The details in the feature selection for the later three nonlinear models can be found in Appendix 3. We also conduct exhaustive hyper-parameter tuning, searching over roughly the same hyper-parameter sets as those when modeling house prices, while we tend to impose more regularization, such as using larger penalty terms and shallower depths of trees, since there are more predictor variables but less samples now.

Table 5 presents the best $R^2$ values for each kind of model and combination of datasets. In general, all the $R^2$ values are lower than 0.1, suggesting that most variance in house price changes cannot be explained by our features. However, such low $R^2$ values are reasonable because individual housing sale prices contains many unobservable noises, such as the transaction costs, the relationships and negotiations between buyers and sellers, and the contract details. This result is also consistent with related work [49].

We are more interested in the differences in the performances by different datasets. First, modeling with only differences in sale time and price index results in the lowest out-of-sample

$R^2$ values, with the highest as 0.0393 by Gradient Tree Boosting. This suggests that the city-wide general information can only explain a very limited part of individual house price changes. Second, the digital census features can improve the out-of-sample $R^2$ values, against baselines with or without housing characteristics or socioeconomic variables. The best $R^2$ value reaches 0.0833 when using all the potential features and fitting with Gradient Tree Boosting. Third, the improvements are statistically significant in most cases, expect when both housing characteristics and socioeconomic variables are incorporated.

Table 5 Average $R^2$ values for modeling house price changes

| Regression model | Only time and index difference | No digital census | 311 complaint | Crime complaint | Taxi trips | All digital census |
|---|---|---|---|---|---|---|
| **Panel A: Without traditional census** | | | | | | |
| Lasso | 0.0248 (0.0010) | 0.0472 (0.0009) | 0.0725 (0.0009) | 0.0675 (0.0008) | 0.0606 (0.0010) | 0.0790 (0.0008) |
| Neural Network | 0.0434 (0.0011) | 0.0525 (0.0009) | 0.0741 (0.0013) | 0.0683 (0.0011) | 0.0641 (0.0014) | 0.0800 (0.0013) |
| Random Forest | 0.0297 (0.0011) | 0.0528 (0.0012) | 0.0782 (0.0009) | 0.0701 (0.0009) | 0.0695 (0.0011) | 0.0818 (0.0009) |
| Gradient Tree Boosting | 0.0393 (0.0012) | 0.0651 (0.0011) | 0.0751 (0.0008) | 0.0695 (0.0010) | 0.0702 (0.0009) | 0.0809 (0.0010) |
| **Panel B: With traditional census** | | | | | | |
| Lasso | 0.0570 (0.0009) | 0.0730 (0.0008) | 0.0803 (0.0009) | 0.0753 (0.0008) | 0.0769 (0.0008) | 0.0821 (0.0009) |
| Neural Network | 0.0606 (0.0011) | 0.0770 (0.0009) | 0.0808 (0.0012) | 0.0784 (0.0012) | 0.0796 (0.0012) | 0.0827 (0.0017) |
| Random Forest | 0.0493 (0.0012) | 0.0684 (0.0010) | 0.0798 (0.0010) | 0.0751 (0.0010) | 0.0773 (0.0010) | 0.0828 (0.0011) |
| Gradient Tree Boosting | 0.0670 (0.0013) | 0.0796 (0.0008) | 0.0814 (0.0010) | 0.0796 (0.0009) | 0.0814 (0.0009) | 0.0833 (0.0009) |

Notes: (1) All columns incorporate differences in sale times and corresponding repeat sales price index; All columns expect the first incorporate basic housing characteristics, by default. (2) The average $R^2$ values are out-of-sample, estimated by five-fold cross validation (grouped by zipcode) repeated for 20 times. (3) The corresponding standard deviations are reported in the parentheses.

## 5 Discussion and conclusion

The objective of this paper is to build a practical and generalizable house price model, and examine whether the digital census can reinforce the predictive power. We select three typical digital census datasets—311 complaints, crime complaints and taxi-trips, and provide comprehensive evidence that they can substantially improve the modeling performances on both house price levels and changes, regardless whether traditional census is included or not. All our models are implemented at the most granular level—individual houses, and geographically grouped cross validations are employed to ensure the modeling results can generalize geographically.

We note that the improvements against the baseline with traditional census are typically smaller than those without traditional census, which suggests that the useful socioeconomic information from digital and traditional censuses is partially overlapped. Therefore, digital census can serve as an effective alternative to traditional census when the latter is absent, especially for real-time applications; digital census can also provide additional information to traditional census, leading to a more comprehensive modeling.

# Appendix 1 Description of housing sales

The housing sales dataset provides the target variable—sale price per unit, and basic predictor variables—housing characteristics for this study. Figure 1 depicts the spatial distribution of sale price per unit across NYC, and Table 6 presents the variables and their descriptive statistics.

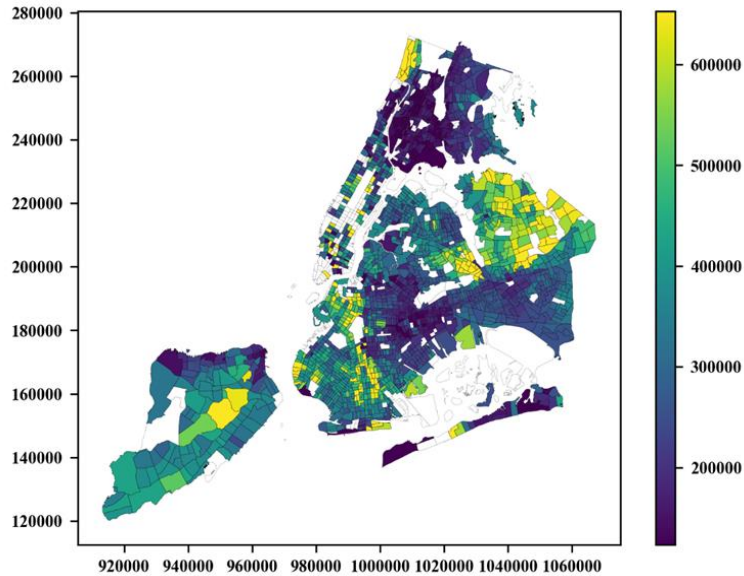

Figure 1 Median sale price per unit for census tracts (US dollar)

Table 6 Descriptive statistics for housing sales data

| Variable | Mean | Std. Dev. |
|---|---:|---:|
| **Sale price per unit (US dollar)** | 315645.2 | 206984.3 |
| **Unit number in the transaction** | 3.372 | 9.919 |
| **Building area per unit (sq. ft.)** | 1076.8 | 420.5 |
| **Floor area ratio (building area divided by lot area)** | 1.058 | 0.783 |
| **Building age (100s years)** | 0.783 | 0.289 |
| **Building number on the lot** | 2.237 | 0.836 |
| **Floor number** | 2.362 | 0.982 |
| **Major alter before sale** | 0.025 | 0.157 |
| **Extension** | 0.098 | 0.298 |
| **Garage** | 0.361 | 0.480 |
| **Full basement** | 0.871 | 0.335 |
| **Partial basement** | 0.006 | 0.076 |
| **Commercial activity** | 0.002 | 0.049 |
| **Building class** | | |
|    Single-family detached house | 0.186 | 0.389 |
|    Single-family attached house | 0.160 | 0.366 |
|    Two-family house | 0.411 | 0.492 |
|    Three-family house | 0.128 | 0.334 |
|    Four-family house | 0.029 | 0.166 |
| Five- or six-family house | 0.029 | 0.168 |
| Over six-family house | 0.033 | 0.179 |
| Walk-up apartment | 0.010 | 0.098 |
| Elevator apartment | 0.012 | 0.108 |
| Multi-use, family house | 0.000 | 0.014 |
| **Lot shape and location** | | |
| Irregular shape | 0.095 | 0.293 |
| Waterfront | 0.001 | 0.023 |
| Corner | 0.075 | 0.263 |
| Through | 0.001 | 0.036 |
| Interior | 0.002 | 0.049 |
| **Distance to city hall (km)** | 45159.4 | 17458.5 |
| **Distance to nearest subway station (km)** | 3737.6 | 4185.1 |
| **Distance to nearest park (km)** | 1744.7 | 1186.4 |

Note: 83,876 observations.

## Appendix 2 Profiles of digital census

We conduct some explorative analysis on the digital census features, providing some basic profiles. We show that the total volumes, weekly timelines and complaint type structures show different patterns across NYC.

Figure 2 visualizes the densities of 311 complaints, crime complaints, taxi pick-ups and drop-offs for each census tract. Manhattan area appears to show highest densities over all datasets, while the local spatial patterns also substantially differ from each other.

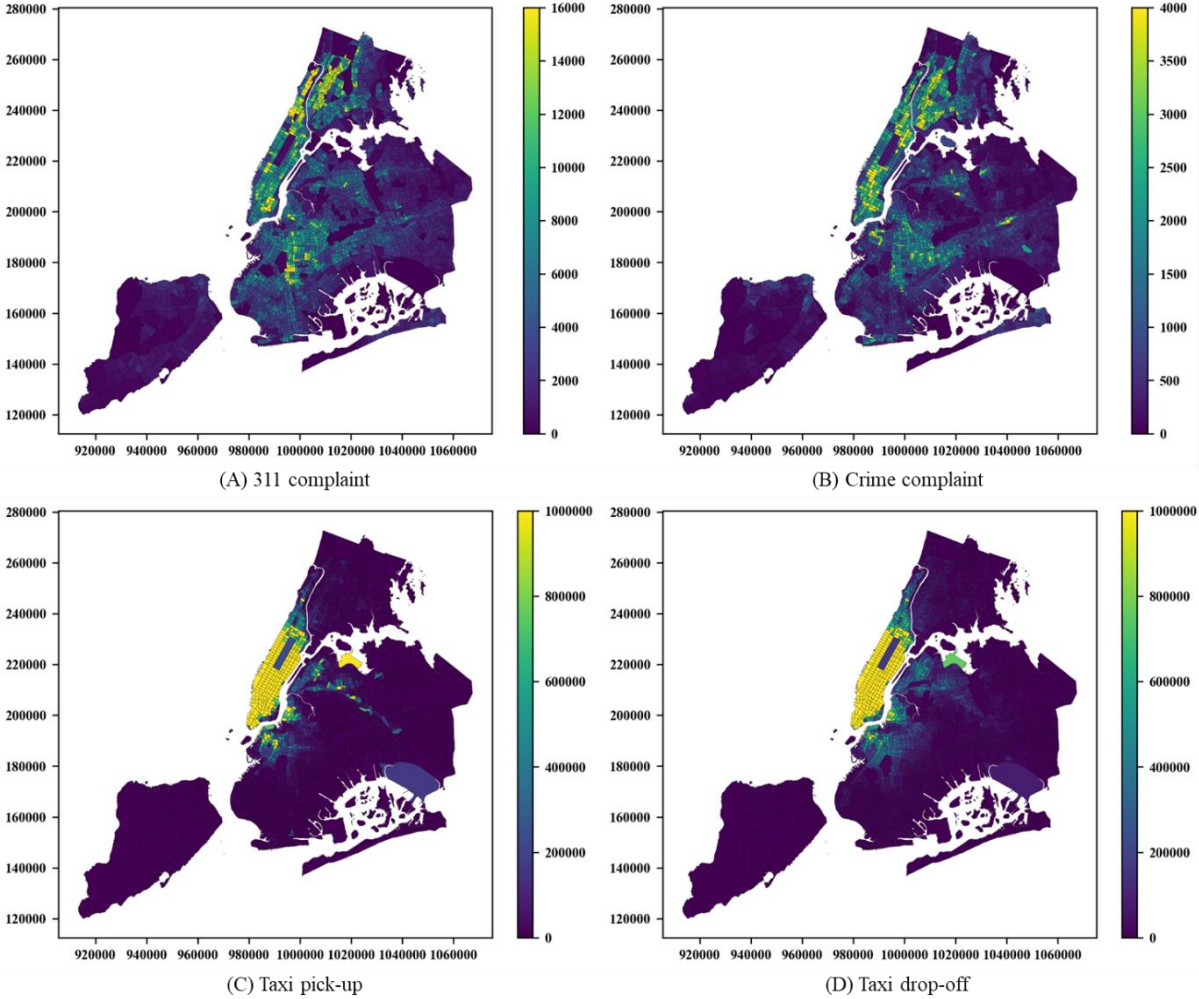

Figure 2 Aggregated digital census volume density for census tracts

Figure 3 presents the normalized weekly timelines of 311 complaints, crime complaints, taxi pick-ups and drop-offs for the five boroughs in NYC. All the weekly timelines are clearly cyclical, with distinct peaks around middays, reflecting the cyclical activity intensities during the typical week. In addition, the timelines of five boroughs show distinctive patterns. For example, Brooklyn shows higher activities of taxi pick-ups than other areas at Friday and Saturday nights, while Staten Island shows higher taxi drop-offs in weekday nights.

Figure 4 presents the top 10 types distribution of 311 and crime complaints for the five boroughs. The complaint type distribution also substantially varies over different areas. For example, Staten Island shows higher relative frequencies of 311 complaints on "street condition", and Manhattan shows higher relative frequencies of crime complaints on "petit larceny" and "grand larceny".

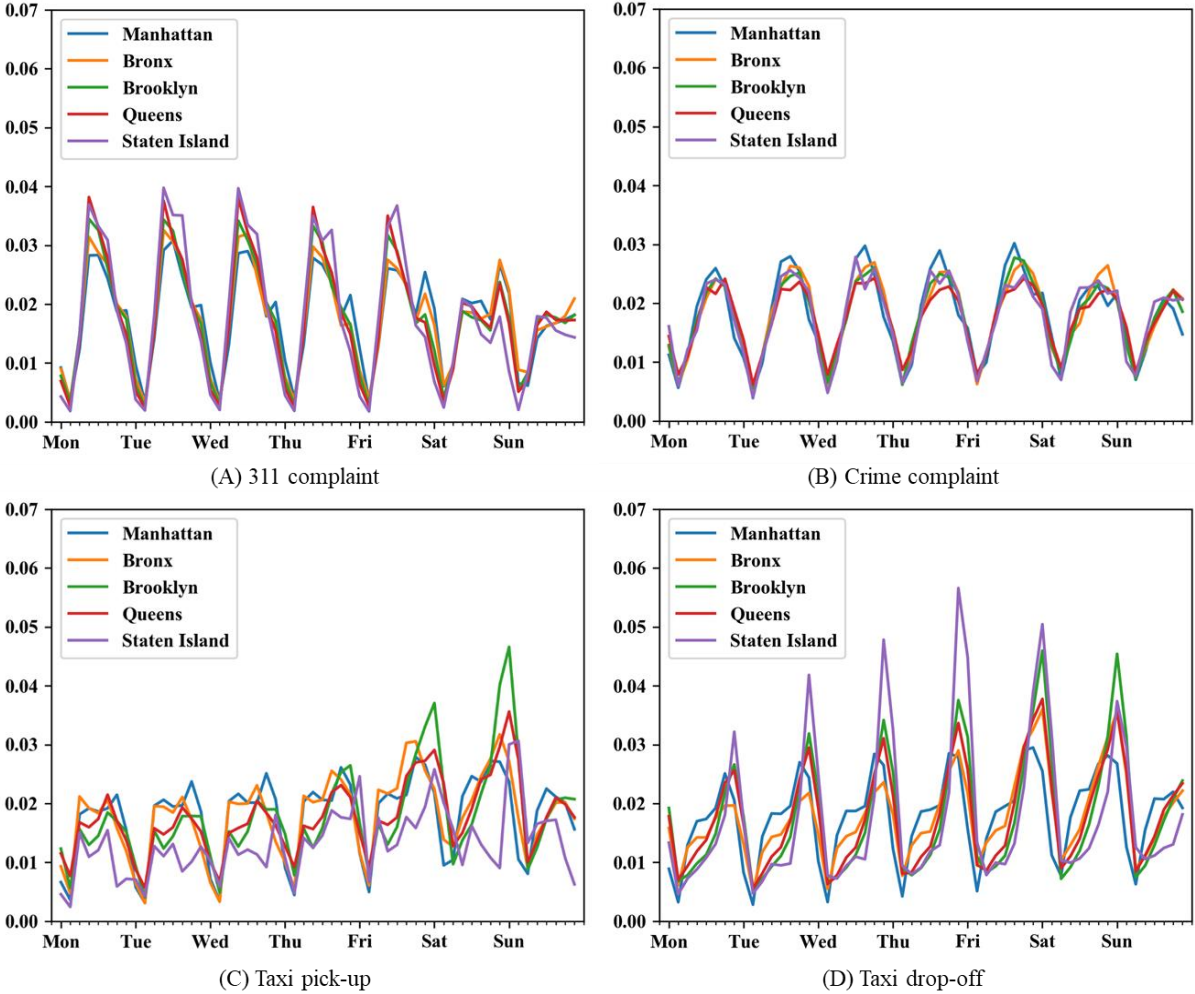

Figure 3 Normalized weekly timelines for five boroughs

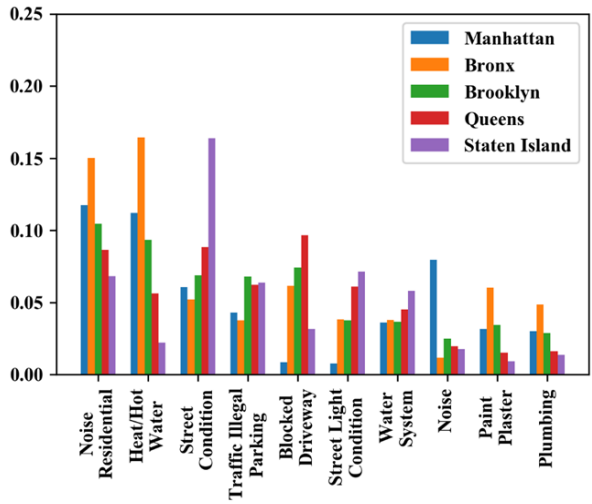 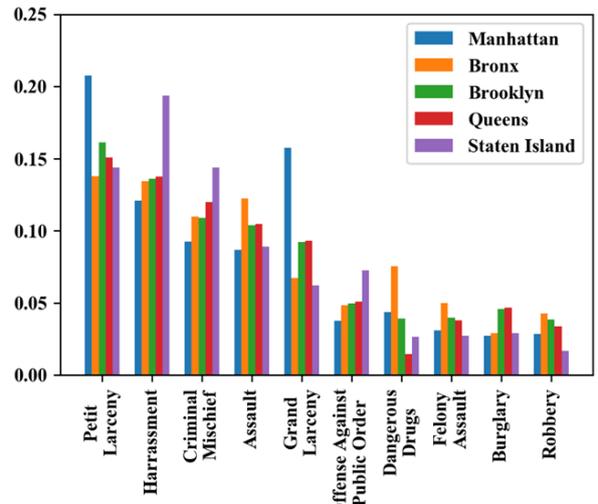

(A) 311 complaint  (B) Crime complaint

Figure 4 Top 10 complaint types distribution for five boroughs

# Appendix 3 Feature selection with Lasso

Before training nonlinear models, we conduct feature selection by fitting Lasso and selecting the features with nonzero coefficients. Figure 5 plots the remaining feature numbers after feature selection by Lasso with different parameter $\alpha$ (L1-penalty) imposed. The remaining feature numbers of all categories decrease with the increase of parameter $\alpha$.

As prior knowledge, the basic housing characteristics and socioeconomic variables have important effects on house prices, so most of them should remain after feature selection. On the other hand, for the purpose of feature selection, the uninformative digital census features should be adequately excluded. According to the figure, we choose to specify the parameter $\alpha$ ranging from 1e-4.5 to 1e-3.5.

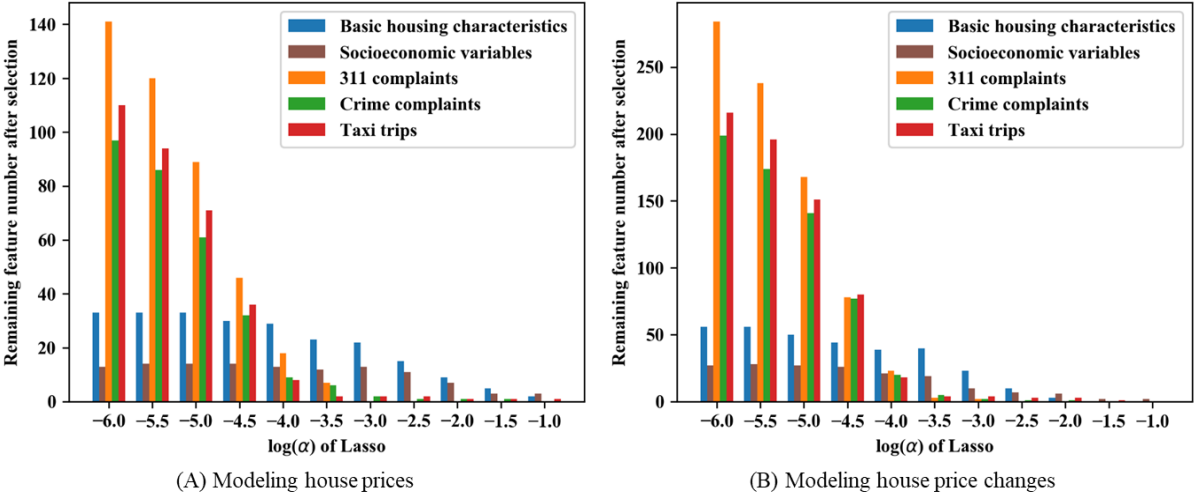

Figure 5 Remaining feature numbers after selection by Lasso